%% file: camera_ready.tex
\documentclass[conference]{IEEEtran}
\IEEEoverridecommandlockouts
\usepackage{cite}
\usepackage{array}
\usepackage{booktabs} 
\usepackage{xcolor, colortbl}
\usepackage{graphicx}
\usepackage{float}
\usepackage{xurl}
\usepackage{amsmath}
\usepackage{subcaption}
\usepackage{tikz}
\usetikzlibrary{shapes.geometric, arrows.meta, positioning, fit, backgrounds}
\usepackage{booktabs} 
\usepackage{tabularx} 
\usepackage{multirow} 
\AtBeginDocument{%
  }

\begin{document}

\title{Quantifying the Blockchain Trilemma: A Comparative Analysis of Algorand, Ethereum 2.0, and Beyond}

\author{
  \IEEEauthorblockN{%
    Yihang Fu\IEEEauthorrefmark{2}
    Mingwei Jing\IEEEauthorrefmark{3}
    Jiaolun Zhou\IEEEauthorrefmark{2}
    Peilin Wu\IEEEauthorrefmark{2}
    Ye Wang\IEEEauthorrefmark{4}
    Luyao Zhang\IEEEauthorrefmark{2}\textsuperscript{*} 
    Chuang Hu\IEEEauthorrefmark{3}\textsuperscript{*} \thanks{\textsuperscript{*} Corresponding authors: Luyao Zhang (lz183@duke.edu), Data Science Research Center and Social Science Division, Duke Kunshan University (DKU) and Chuang Hu (handc@whu.edu.cn), Department of Computer Science, Wuhan University (WHU). \textbf{Acknowledgments.} The research results of this article (or publication) are sponsored by the Kunshan Municipal Government research funding for the DKU-WHU Joint Seeding Project entitled "Computational Economics."}
  }
  \IEEEauthorblockA{%
    \IEEEauthorrefmark{2}Duke Kunshan University, 8 Duke Ave., Suzhou, China, 215316\\
    \IEEEauthorrefmark{3} Wuhan Unversity, Wuchang District, Wuhan, Hubei, China, 430072\\
    \IEEEauthorrefmark{4} University of Macau, Avenida da Universidade, Taipa, Macau, China.
  }
}

\maketitle

\begin{abstract}
Blockchain technology is essential for the digital economy and metaverse, supporting applications from decentralized finance to virtual assets. However, its potential is constrained by the "Blockchain Trilemma," which necessitates balancing decentralization, security, and scalability. This study evaluates and compares two leading proof-of-stake (PoS) systems, Algorand and Ethereum 2.0, against these critical metrics. Our research interprets existing indices to measure decentralization, evaluates scalability through transactional data, and assesses security by identifying potential vulnerabilities. Utilizing real-world data, we analyze each platform's strategies in a structured manner to understand their effectiveness in addressing trilemma challenges. The findings highlight each platform's strengths and propose general methodologies for evaluating key blockchain characteristics applicable to other systems. This research advances the understanding of blockchain technologies and their implications for the future digital economy. Data and code are available on GitHub as open source.
\end{abstract}



%
\IEEEpeerreviewmaketitle

\section{Introduction}

Blockchain technology has made significant strides, positioning itself as a decentralized framework pivotal for enhancing distributed artificial intelligence~\cite{cao2022decentralized}. However, its evolution is challenged by the "Blockchain Trilemma," which requires a delicate balance among decentralization, scalability, and security~\cite{abadi2018blockchain}. Previous research predominantly focused on earlier blockchain versions, often lacking comparative analyses of performance metrics within more advanced, consistent frameworks~\cite{security_survey}.

Our study addresses this deficiency by evaluating and comparing decentralization, scalability, and security across two proof-of-stake (PoS) systems: Algorand and Ethereum 2.0~\cite{chen2019algorand},~\cite{grandjean2023ethereum}. We explore essential questions below:

\begin{itemize}
    \item How to interpret existing indices that Measure the Decentralization of Algorand and Ethereum 2.0?
    \item How to Measure the Scalability of Algorand and Ethereum 2.0?
    \item How to Measure the Security of Algorand and Ethereum 2.0?
\end{itemize}

Utilizing real-world data, our analysis examines each platform's approach to these metrics in a scientific and structured manner. We investigate decentralization using established indices, evaluate scalability through transactional data, and assess security by identifying potential vulnerabilities and their defenses.

\input{fig/fig0}

Figure~\ref{fig:flowchart} outlines the structure of our paper. Section~\ref{sec:related work} introduces related work, while Section~\ref{sec:method} describes our methodology. The findings are presented in Section~\ref{sec:results}, followed by a discussion of the implications in Section~\ref{sec:discussion}. Section~\ref{sec: future} points out the limitations and highlights future research directions.

This research provides a comparative analysis of Algorand and Ethereum 2.0 while also developing general methodologies for evaluating essential blockchain characteristics. These methodologies are designed to be applicable to other blockchain systems, thereby advancing our understanding of blockchain technologies and their implications for the future digital economy.

\textbf{Data and Code Availability Statements}. The on-chain data used in the paper is provided in Appendix Table \ref{table:5}, and both the data and code are available on GitHub at \url{https://github.com/KerwinFuyihang/blockchain_analysis}.

\section{Related Work}
\label{sec:related work}
In this section, we begin with a concise introduction to Algorand and Ethereum 2.0. Subsequently, we provide a summary of existing research on blockchain metrics.

\subsection{Algorand and Ethereum 2.0}
The Proof-of-Stake (PoS) protocol has emerged as a more efficient and environmentally friendly alternative to the traditional Proof-of-Work (PoW) protocol~\cite{zhang2023understand}. To fully understand and evaluate PoS-based systems, it is essential to establish reliable methods for assessing their critical metrics. While PoS serves as the foundation for several blockchain systems, its implementations vary. Among these, Algorand and Ethereum 2.0 are notable examples.

Algorand introduces an innovative consensus algorithm that integrates PoS with the Verifiable Random Function (VRF)~\cite{vrf}, enabling all participants to stake their tokens and actively engage in the blockchain's operations. In contrast, Ethereum 2.0 employs a PoS-based consensus mechanism where participants must stake a specific amount of tokens to earn validation rights, following a series of verifications. 

Ethereum 2.0 features two distinct layers: the consensus layer and the execution layer. The consensus layer, formerly known as the Beacon Chain, was established to transition Ethereum from a PoW to a PoS consensus mechanism. The execution layer, a continuation of the original Ethereum blockchain (Eth1), is responsible for transaction processing and smart contract execution. It works in tandem with the consensus layer, which coordinates validators and manages consensus across the network~\cite{eth2000}.

By comparing the metrics of Algorand and Ethereum 2.0, we can gain deeper insights into the PoS mechanism, enhancing our ability to accurately quantify blockchain metrics.

\subsection{Decentralization} 
Traditionally, decentralization is defined as the absence of central coordination. In the context of blockchain, it refers to the distribution of control and decision-making across the network, eliminating the need for a central authority~\cite{Zhang23}. This distribution enhances the system's transparency, security, and resilience by preventing any single entity from holding control~\cite{zhang2023design}.

Existing research indicates that blockchain decentralization is controversial~\cite{xiao2024centralized, wu2024trust} and multifaceted, encompassing dimensions such as hardware, software, network, consensus, and transactions~\cite{karakostas2022sok}. Recent studies~\cite{gochhayat2020measuring, lin2021measuring, zhang2023sok, zhang2023blockchain, ao2022decentralized, chemaya2023uniswap} have introduced various mathematical methods to quantify these aspects of decentralization. These methods employ coefficients such as Shannon Entropy, Gini Coefficient, Nakamoto Coefficient, and the Herfindahl-Hirschman Index, along with network features in relevant case studies, to measure decentralization effectively.

\subsection{Scalability}
Scalability is a critical aspect of blockchain research, primarily concerned with the overall efficiency of blockchain systems. Enhanced scalability implies reduced resource costs in blockchain transactions~\cite{mccorry2021sok}. A case study on Bitcoin~\cite{scalability_1} introduces a set of metrics to evaluate scalability, including maximum throughput, latency, and transaction throughput. Further research~\cite{scalability_2} identifies maximum throughput and cost as key components for quantifying blockchain scalability.

\subsection{Security}
Security is a fundamental property of a blockchain system, deriving from its nature as a distributed ledger that emphasizes reliability and integrity. Conventional security issues in blockchain can be categorized into several types, including 51\% attacks, forking issues, and eclipse attacks, among others~\cite{security_review}. Further exploration reveals that blockchain security issues are complex and can be subdivided based on their causes, such as operational mechanisms and smart contracts~\cite{security_survey}. 

Despite extensive research, there remains a lack of comprehensive methods for evaluating blockchain security. Current studies predominantly focus on enhancing security techniques in response to real-world attacks, such as the infamous "DAO" attack. Therefore, there is a critical need for developing efficient methods to evaluate the security capacity of blockchain systems to preemptively address potential threats.

\section{Methodology}
\label{sec:method}
This section presents our methodology for evaluating Algorand and Ethereum 2.0, focusing on the key aspects of decentralization, scalability, and security using real-world data collected from BitQuery and Beacon Explorer.

\subsection{Data Description}
\input{tabs/tab1}

\input{tabs/tab2}
We collected on-chain data for Algorand from January 2019 to September 2023 via BitQuery's open APIs, and for Ethereum 2.0 from June 2019 to September 2023 through Beacon Explorer using the SPIDER framework. Our data encompasses blocks, transactions, and accounts, which we categorize according to the targeted metrics detailed in Tables \ref{table:1} and \ref{table:2}. We also provide a more detailed data dictionary in Table~\ref{tab: dictionary}.
\subsection{Empirical Analysis}
Our analysis assesses:
\begin{itemize}
    \item \textbf{Decentralization:} We explore decentralization at the consensus and transaction layers, employing metrics like the Shannon Entropy, Gini Coefficient, Nakamoto Coefficient, and Herfindahl Hirschman Index. The decentralization indices are defined in Appendix \ref{subsec: indices}.
    \item \textbf{Scalability:} Scalability is analyzed in terms of throughput—transactions per second—and latency—time to confirm transactions. This evaluation uses comparative data to identify performance under normal and peak loads.
    \item \textbf{Security:} Security analysis is split into:
        \begin{itemize}
            \item \textbf{Real Data Analysis:} Examining the correlation between burned fees and security incentives, we propose that higher fees could signify a more secure network.
            \item \textbf{Theoretical Comparison:} Assessing each platform's vulnerability to attacks, particularly focusing on mechanisms like Algorand’s \textbf{Verifiable Random Function} (VRF) ~\cite{vrf} and Ethereum 2.0’s RANDAO~\cite{rando_specs}\footnote{\url{https://github.com/randao/randao}}, and conducting an empirical test scenario of a 51\% attack.
        \end{itemize}
\end{itemize}

This structured approach allows us to address the complexities of the Blockchain Trilemma through a comprehensive examination of each platform.

\section{Results}
\label{sec:results}
In this section, we present our empirical results and conduct a comprehensive analysis to reveal the insights obtained from our empirical evaluations.

\subsection{R1: Mixed Features of Decentralization}
\begin{figure}[!htbp]
    \centering
    \centering
    \begin{subfigure}[b]{0.47\textwidth}
        \centering
        \includegraphics[width=1.0\textwidth]{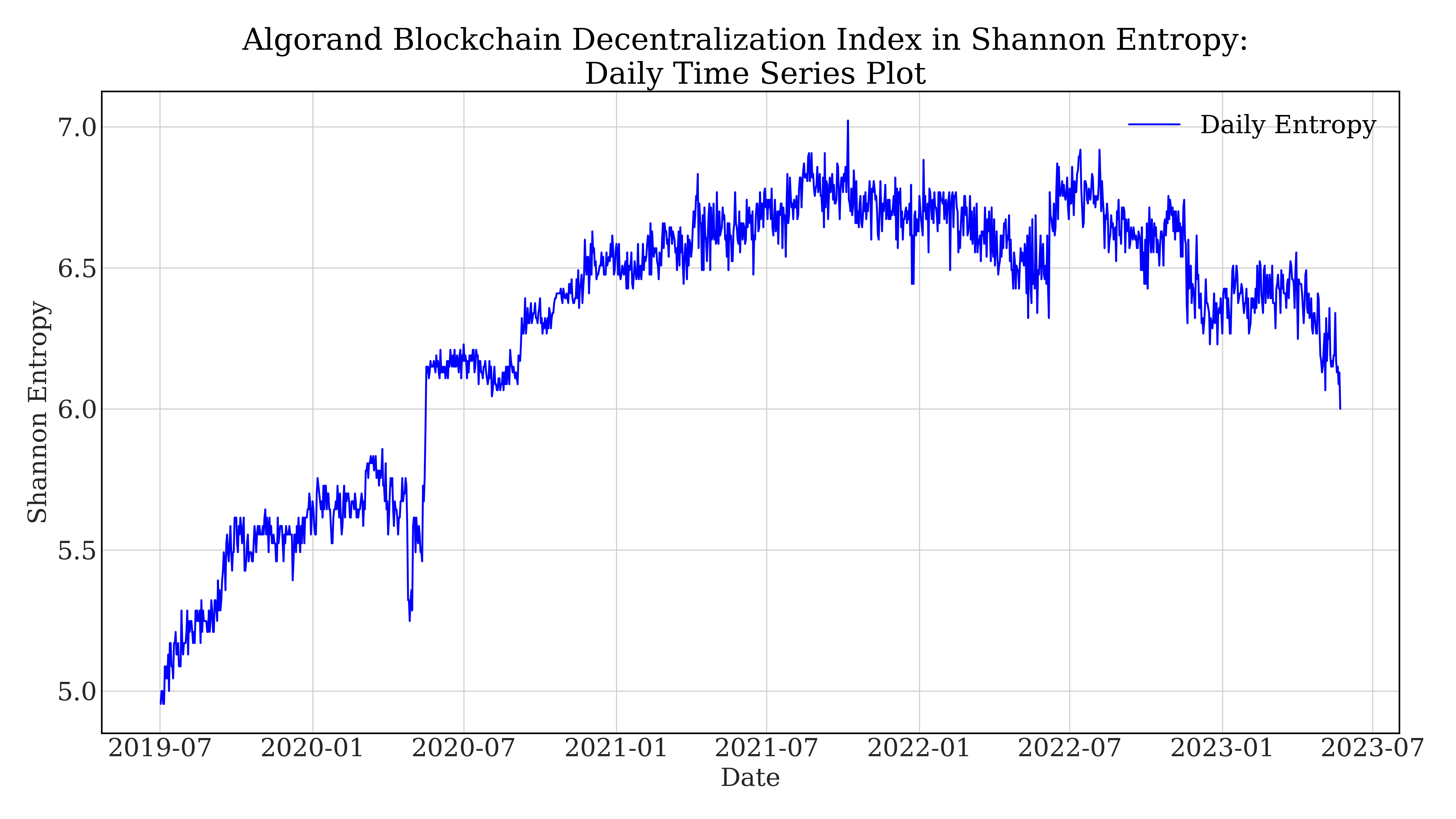}
        \caption{Daily Shannon Entropy of Algorand on Consensus layer}
        \label{fig:sub1}
    \end{subfigure}
    \hfill
    \begin{subfigure}[b]{0.47\textwidth}
        \centering
        \includegraphics[width=1\textwidth]{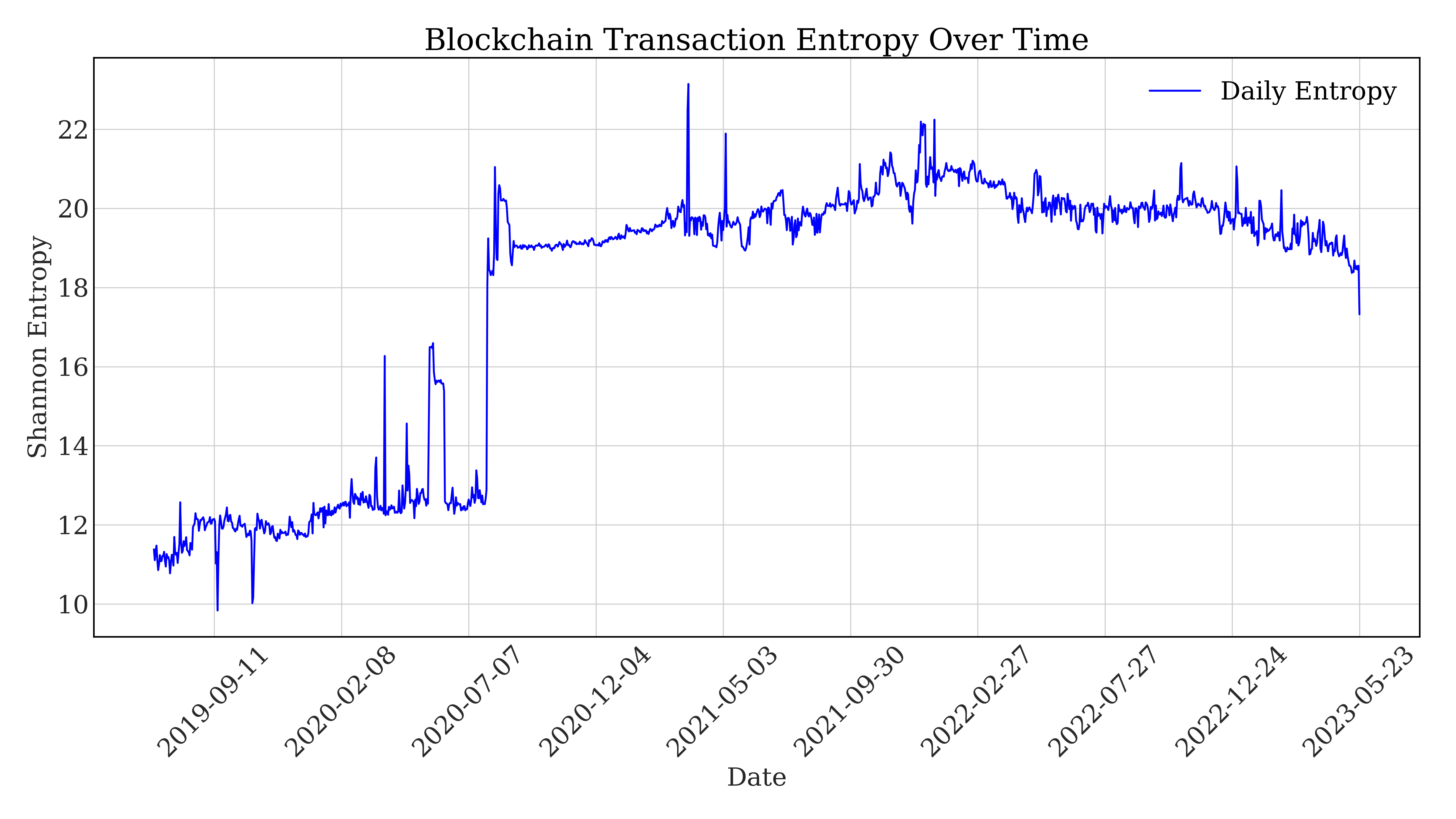}
        \caption{Daily Shannon Entropy of Algorand on Transaction layer}
        \label{fig:sub2}
    \end{subfigure}

    \begin{subfigure}[b]{0.47\textwidth}
        \centering
        \includegraphics[width=1\textwidth]{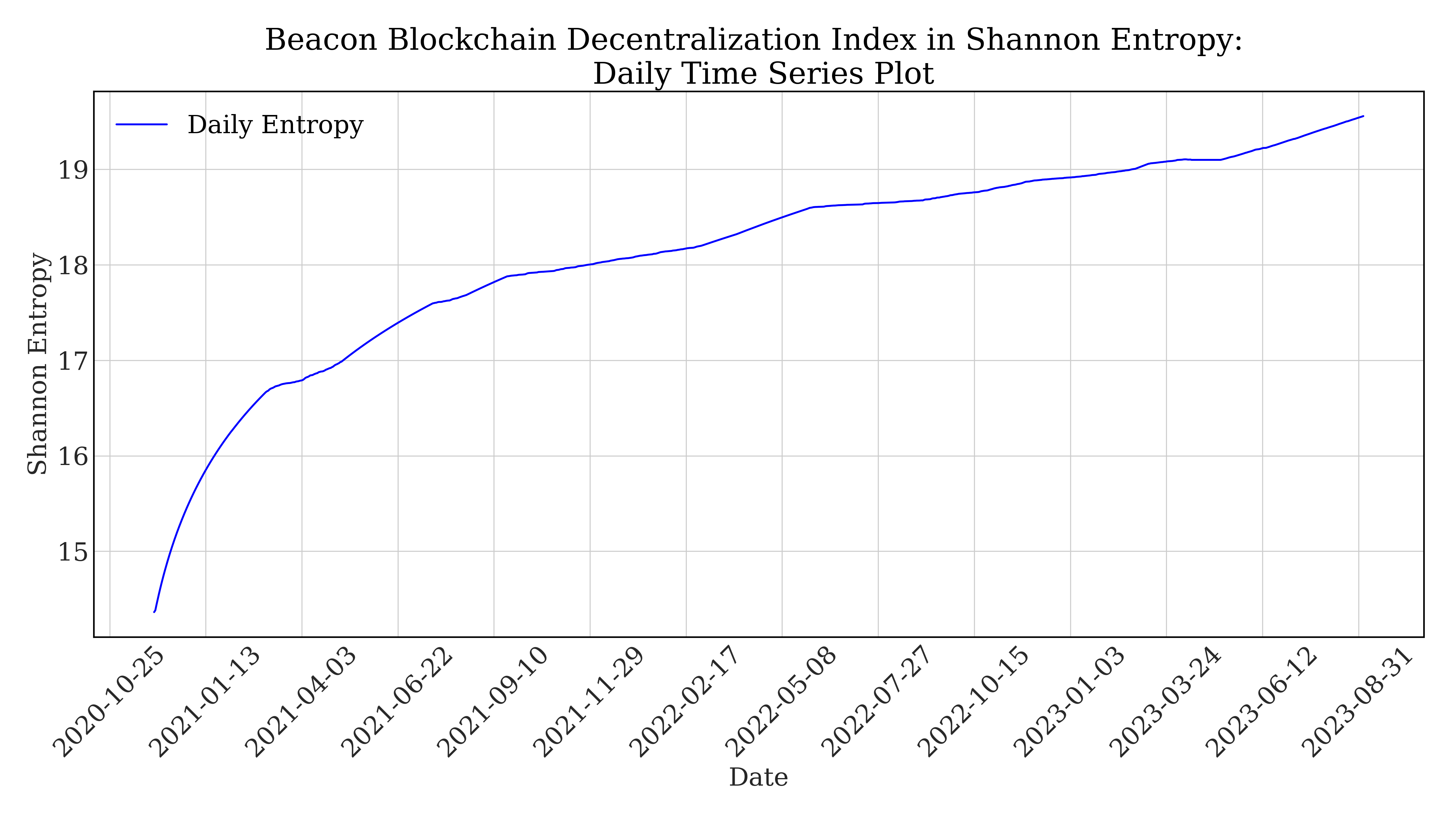}
        \caption{Daily Shannon Entropy of Ethereum 2.0 on Consensus layer}
        \label{fig:sub3}
    \end{subfigure}
    \hfill
    \begin{subfigure}[b]{0.47\textwidth}
        \centering
        \includegraphics[width=1\textwidth]{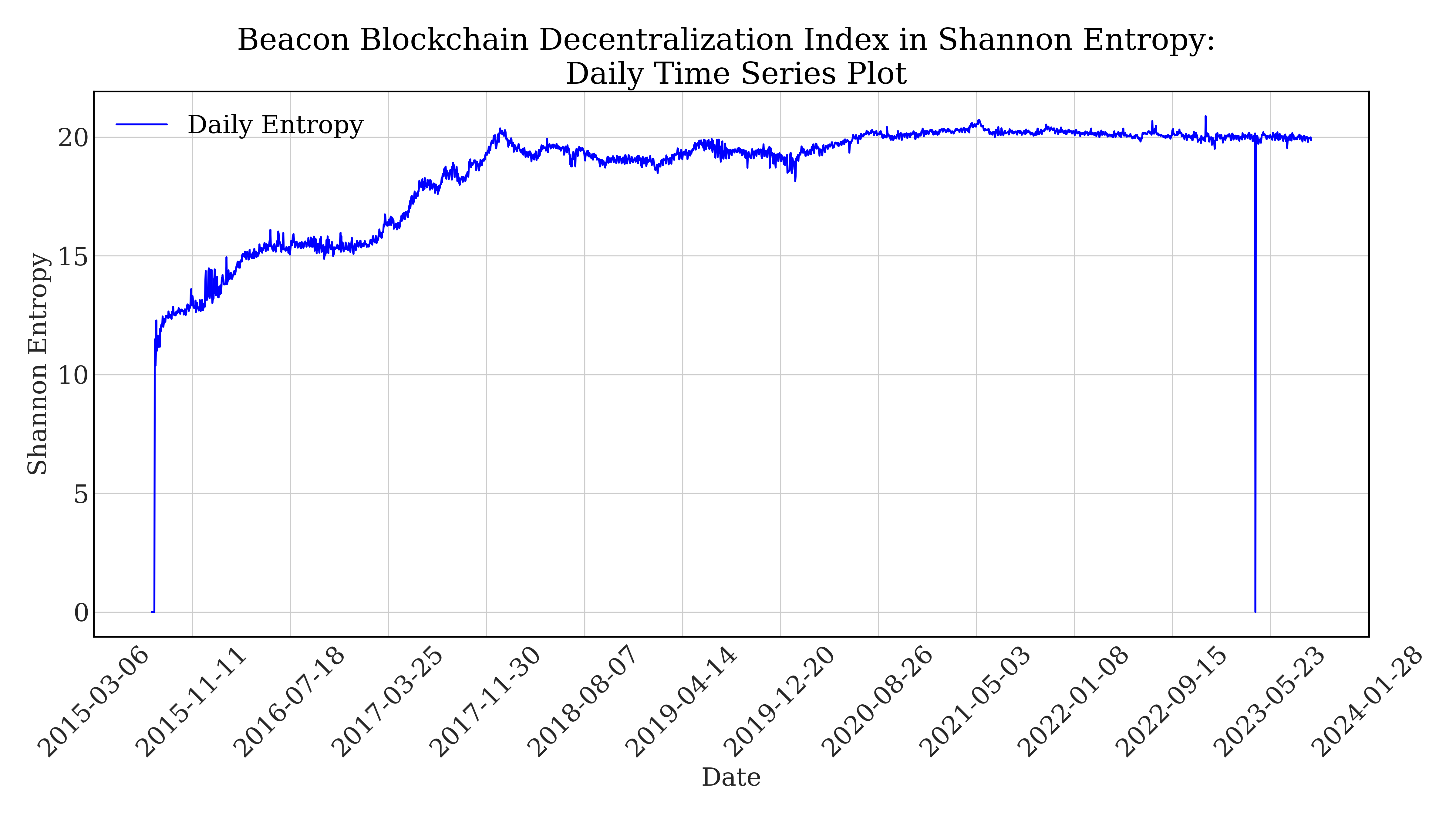}
        \caption{Daily Shannon Entropy of Ethereum 2.0 on Transaction layer}
        \label{fig:sub4}
    \end{subfigure}
    
    \caption{Daily Shannon Entropy of Algorand and Ethereum 2.0 on both consensus and transaction layer.}
    \label{fig:test}
\end{figure}

We first present the results of Shannon entropy applied to the consensus and transaction layers of Algorand and Ethereum 2.0, as illustrated in Figure~\ref{fig:test}.

Figure~\ref{fig:test} illustrates an apparent heterogeneity between the decentralization of the consensus layer and transaction layers' decentralization. Despite fluctuations, the trends suggest an overall increase in decentralization over time, with a significant peak in the Algorand consensus layer around 300 days after the initial date recorded.

\input{tabs/tab3}

\begin{figure}[!h]
    \centering
    \begin{subfigure}[b]{0.49\textwidth}
        \centering
        \includegraphics[width=\textwidth]{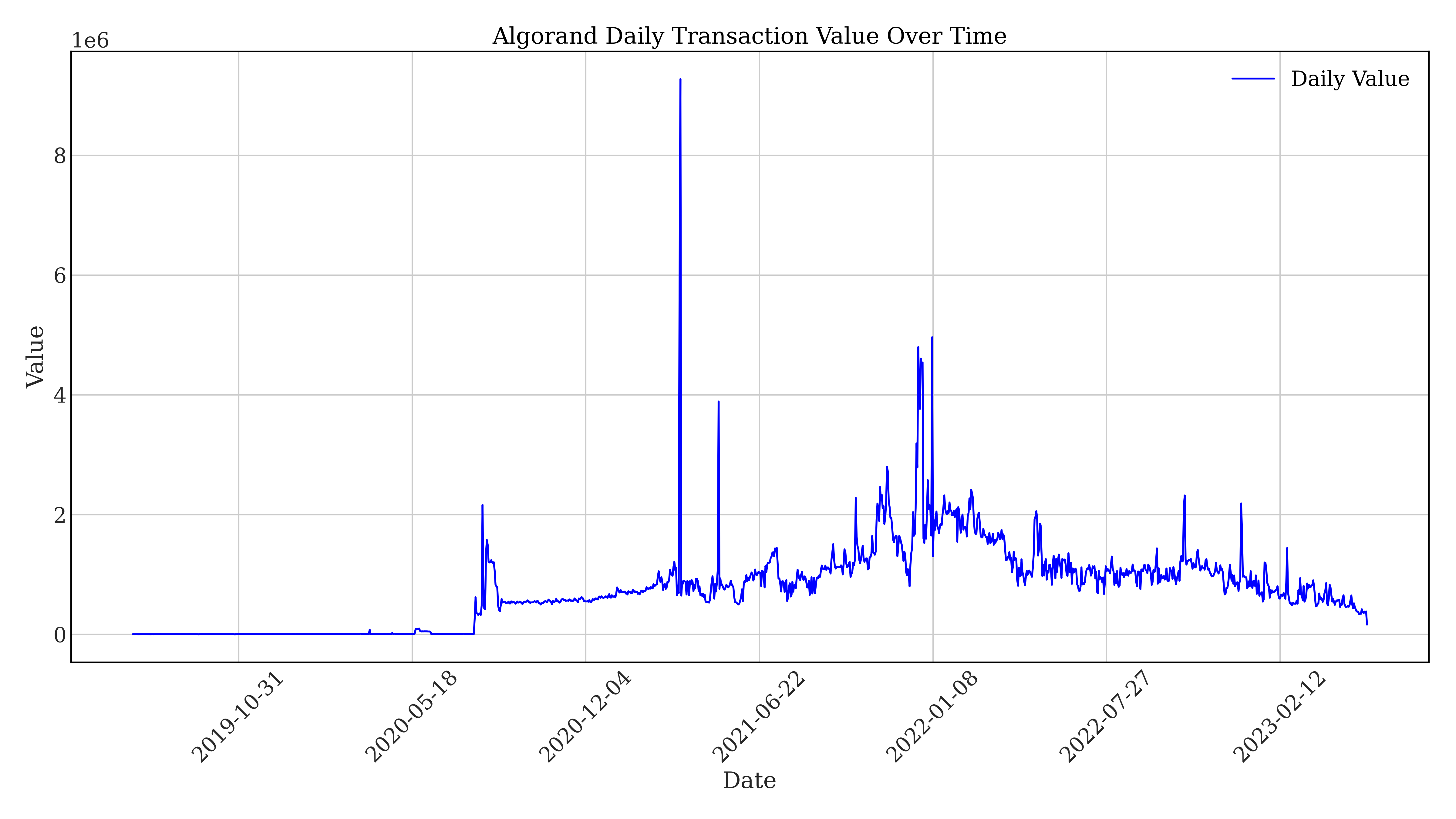}
        \caption{Algorand Daily Transaction}
        \label{fig:algorand_trans01}
    \end{subfigure}
    \hfill
    \begin{subfigure}[b]{0.49\textwidth}
        \centering
        \includegraphics[width=\textwidth]{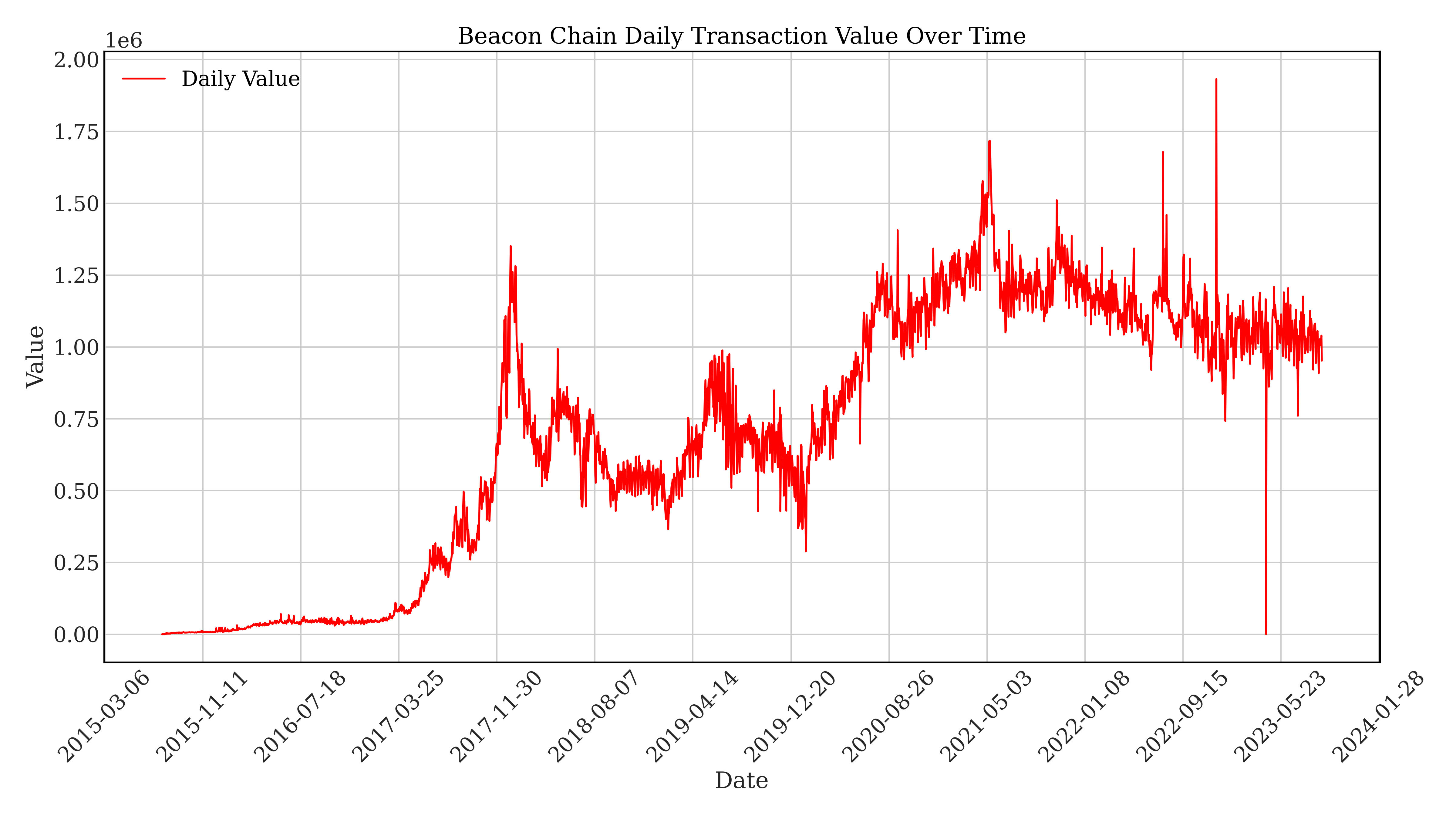}
        \caption{Beacon Chain Daily Transaction}
        \label{fig:beacon_trans01}
    \end{subfigure}
    \caption{The Daily Transaction Data of Ethereum 2.0 and Algorand}
    \label{fig:trans}
\end{figure}

Table~\ref{table:dec_result} details the computed decentralization indices over time. \textbf{On the consensus layer, Algorand exhibits greater decentralization than Ethereum 2.0}, evidenced by higher Shannon Entropy and Nakamoto Coefficient and lower Gini Coefficient and Herfindahl Hirschman Index (HHI). This aligns with Algorand's design to mitigate the "blockchain trilemma." Unlike Ethereum 2.0, which requires a token stake for participation, Algorand's mechanism allows open participation in the voting processes, enhancing its flexibility.
Conversely, the transaction layer shows contrasting trends. While Shannon Entropy and Nakamoto Coefficient suggest greater decentralization for Ethereum 2.0, the Gini Coefficient and HHI favor Algorand. Ethereum's longer operational history and higher transaction volume likely contribute to a more even distribution. Algorand, with its shorter history and less consistent transaction volumes, exhibits peaks of activity that suggest a less uniform distribution, as evidenced in Figure~\ref{fig:trans}.

\subsection{R2: Algorand Gains More Scalability}
Figure \ref{fig:trans} illustrates the transaction throughput of Ethereum 2.0 compared to Algorand. It is evident that the overall transaction volume of Ethereum 2.0 substantially exceeds that of Algorand, which is expected given Ethereum 2.0's popularity in the cryptocurrency market. Surprisingly, the peak transaction volume of Algorand surpasses that of Ethereum 2.0, suggesting that despite Ethereum 2.0's greater popularity and perceived reliability, Algorand may handle more transactions under extreme conditions.

\begin{figure}[!htbp]
    \centering
    \includegraphics[width=0.47\textwidth]{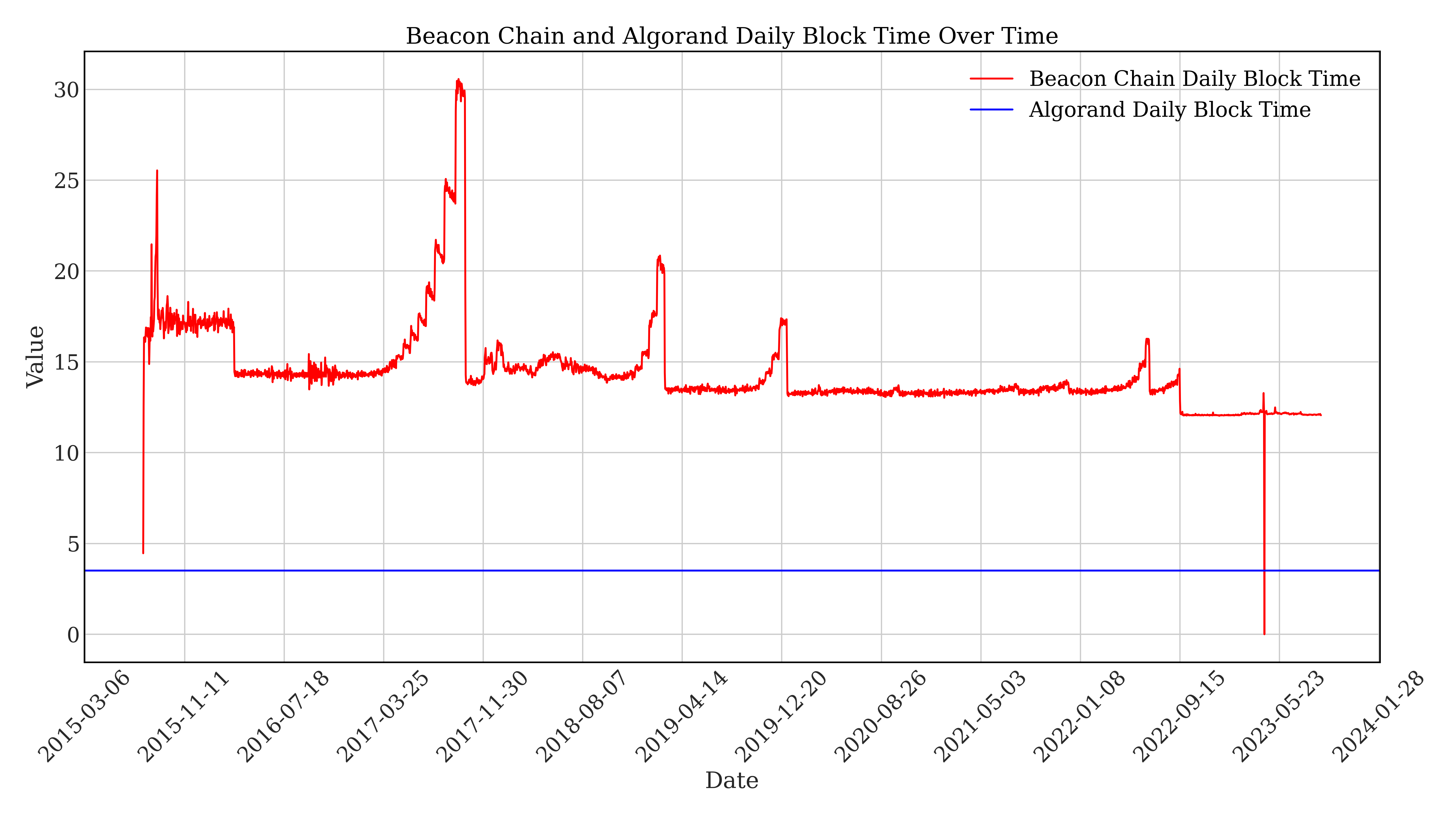}
    \caption{The Daily Block Time of Ethereum 2.0}
    \label{fig/Beacon_Algorand_daily_block_time.png}
\end{figure}

Figure \ref{fig/Beacon_Algorand_daily_block_time.png} displays the latency behavior of Ethereum 2.0. Since the statistics for Algorand remain constant in our records, they are not included in the graph. Generally, the latency data for Ethereum 2.0 demonstrates a more stable trend compared to its transaction data. Notably, the average block time for Algorand is \textbf{3.5s}, while for Ethereum 2.0, it is \textbf{14.42s}. This indicates that Algorand's average block time and transaction processing are significantly faster than those of Ethereum 2.0, enabling quicker block production and confirmation.

\begin{figure}[!h]
    \centering
    \begin{subfigure}[b]{0.49\textwidth}
        \centering
        \includegraphics[width=\textwidth]{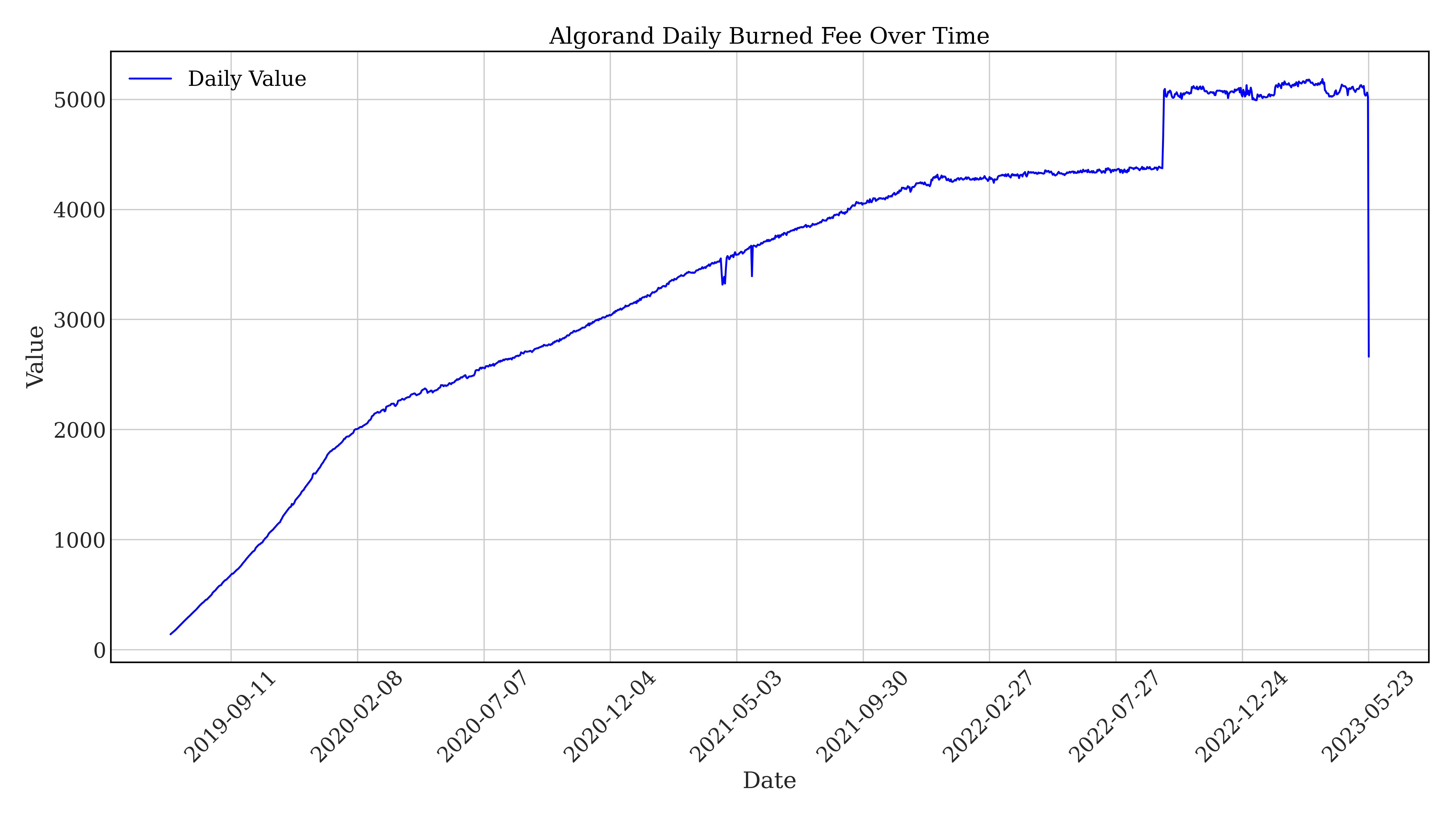}
        \caption{Algorand Daily Burned Fee}
        \label{fig:algorand_trans02}
    \end{subfigure}
    \hfill
    \begin{subfigure}[b]{0.49\textwidth}
        \centering
        \includegraphics[width=\textwidth]{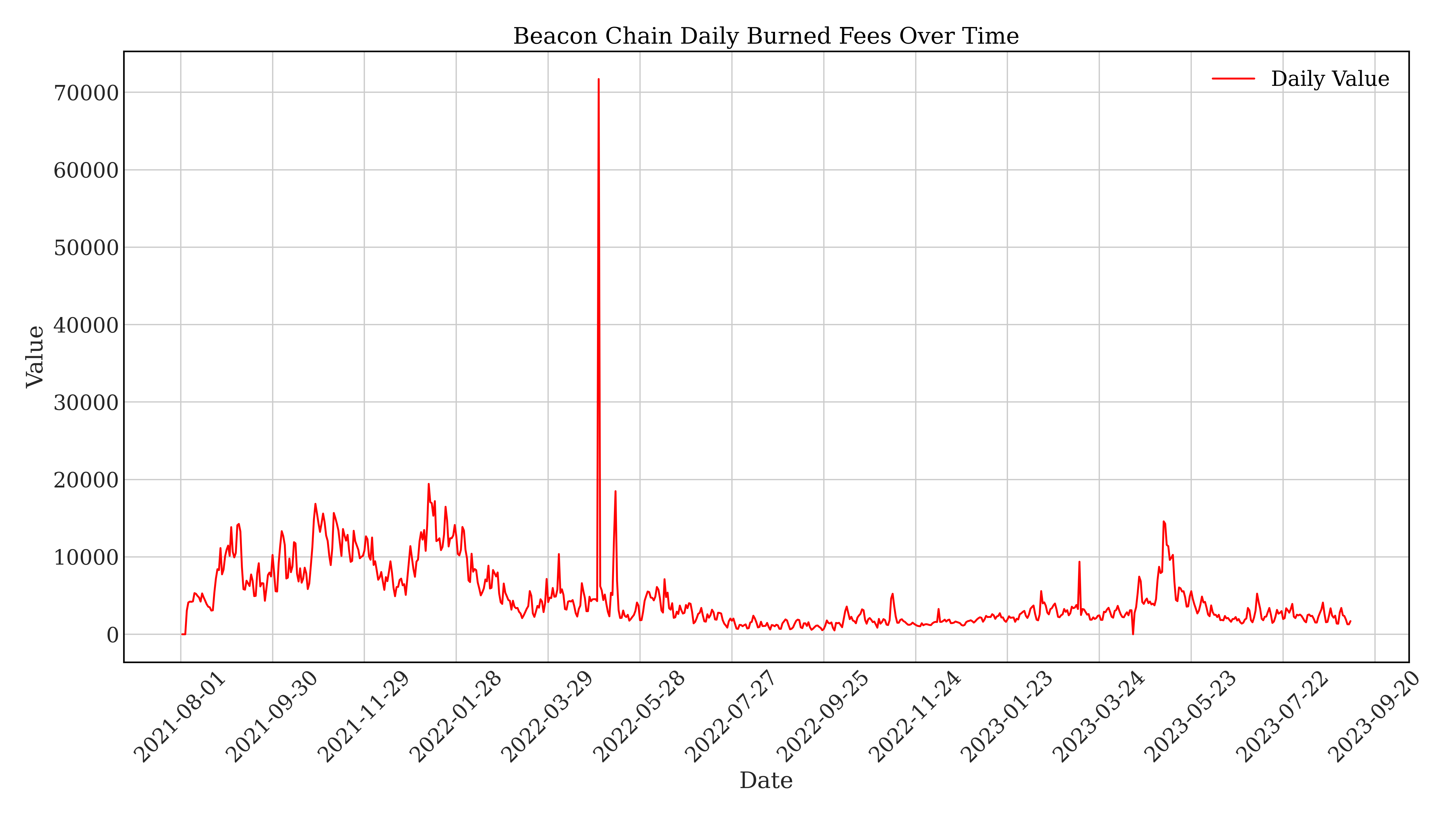}
        \caption{Beacon Chain Daily Burned Fee}
        \label{fig:beacon_trans02}
    \end{subfigure}
    \caption{ The Time Series of Burned Fees of Ethereum 2.0 \& Algorand}
    \label{fig:ts_bu}
\end{figure}
Thus, \textbf{Algorand emerges as more scalable}, achieving higher peak transaction volumes and faster block times. However, the variability in market scale and activity level between Algorand and Ethereum 2.0 introduces uncertainties in our analysis. To obtain more definitive conclusions about their scalability, further in-depth evaluations are needed.

\subsection{R3: Underlying the Secret of Security}
\subsection{Real Data Analysis} 
Security, a critical yet abstract metric, is examined through empirical data analysis. Figure \ref{fig:algorand_trans02} and \ref{fig:beacon_trans02} present the time series of burned fees for Algorand and Ethereum 2.0. The average daily burned fees for Ethereum 2.0 is \textbf{4690.36}, in contrast to Algorand's \textbf{3401.82}, indicating higher transaction costs for Ethereum 2.0. According to the Honest Majority Money (HMM) hypothesis~\cite{algorand_exp}, a system's security is likely guaranteed if the majority remains honest, as they tend to protect the community. The incentive structure, including rewards and minimal inflation, suggests that Ethereum 2.0 could potentially achieve higher long-term security.
\subsection{Theoretical Comparison} 
Given the scarcity of recorded attacks on both Algorand and Ethereum 2.0, we analyze how these platforms would potentially handle the classic 51\% attack~\cite{51attack}. Such attacks involve malicious nodes acquiring control over 51\% of the voting power, allowing them to manipulate the consensus process and, subsequently, the blockchain's behavior.

In the hypothetical scenario where attackers control 51\% of validators or proposers, they could dictate the outcome of proposals. To counteract such risks, enhancing system randomness is crucial, as it prevents attackers from predicting and influencing subsequent block selections. Algorand employs a mechanism known as random seed Q~\cite{algorand_specs}, which updates independently with each voting round, thus ensuring that transaction volumes do not skew randomness. Conversely, Ethereum 2.0 uses a function called RANDO~\cite{rando_specs}, which achieves randomness by amalgamating the current round's random value with the previous one using an XOR operation with timestamps.

Moreover, while a 51\% attack theoretically poses a significant threat, it is rendered impractical in these systems~\cite{algorand_specs,rando_specs}. For Ethereum 2.0, accumulating 51\% of the total stake is considered unfeasible, effectively neutralizing the threat of such an attack. On the other hand, Algorand's validation selection process resembles a lottery, where each validator's chance to participate is temporary and equally likely. This lottery-like mechanism ensures that once a validator's role in consensus is completed, the temporary key is discarded, safeguarding the system against biases and maintaining integrity even under corrupt influences.

In conclusion, both Algorand and Ethereum 2.0 incorporate robust measures to ensure randomness and safeguard against the theoretical possibility of a 51\% attack. However, continuous empirical research is necessary to further validate these defenses under various operational conditions.

In conclusion, both Algorand and Ethereum 2.0 exhibit strong randomization and robustness against theoretical attacks, according to the literature, yet further empirical analysis is necessary for a more definitive comparison.
\section{Discussion and Conclusion}
\label{sec:discussion}
Our empirical analysis reveals that Algorand achieves greater decentralization compared to Ethereum 2.0, reflecting their foundational goals. Algorand supports unrestricted participant engagement in voting and validation, contrasting with Ethereum 2.0’s focus on stability through token staking requirements. This difference is evident in our decentralization metrics and is further underscored by a notable increase in Algorand participants from January to May 2020, likely influenced by Algorand’s \$50 million educational initiative and a strategic bridge to Ethereum, enhancing its DApp ecosystem connectivity.

Algorand’s inclusive design also contributes to superior scalability, indicated by higher transaction volumes and reduced block times, validating our analysis methods. However, security comparisons are less definitive. Preliminary data on burned fees suggest that Ethereum 2.0 may be more secure, encouraging honesty through higher participant costs, though further research is needed for a conclusive assessment.

This study provides a comparative insight into blockchain decentralization, scalability, and security, highlighting Algorand's and Ethereum 2.0’s distinct approaches and outcomes. Despite thorough analysis, the absence of standardized metrics for these core attributes points to a significant research gap. Future work should aim to establish universally recognized benchmarks, potentially through collaborative academic endeavors, to effectively navigate the trade-offs inherent in blockchain development.

\section{Limitations and Future Research Directions}
\label{sec: future}

Blockchain technology is transforming the digital economy by removing intermediaries and enabling the creation of extensive open-source data. This transformation is bolstered by advancements in Layer 2 (L2) solutions, which address the efficiency and scalability limitations of Layer 1 (L1) systems~\cite{chemaya2023uniswap}. Furthermore, blockchain has evolved to incorporate networks of subnets for L2 solutions and meta-networks for cross-chain interoperability~\cite{augusto2023sok}.

The introduction of multinetwork and layered architectures adds complexities and challenges to measuring and assessing blockchain system features. However, it also presents opportunities for integrating blockchain with collaborative machine learning, particularly federated analytics, to enhance the assessment process~\cite{zhang2023machine}.

\textbf{Federated Analytics}: This enables the analysis of data distributed across multiple entities while maintaining data localization, which is invaluable in scenarios where data centralization is impractical due to privacy, regulatory, and bandwidth constraints~\cite{wang2021federated}. The structure of blockchain, characterized by its diverse subnets with specific functions, user groups, and geographic distributions, parallels the distributed nature of federated analytics. This paper proposes conceptualizing federated analytics clients as analogous to blockchain subnets. These subnets could compute a global index while preserving data localization through secure communication protocols and collaborative algorithms, thereby maintaining privacy and leveraging blockchain's inherent security and decentralization to enhance the robustness of federated analytics~\cite{shi2023federated}.

\begin{figure}[!htbp]
\centering
\includegraphics[width=.47\textwidth]{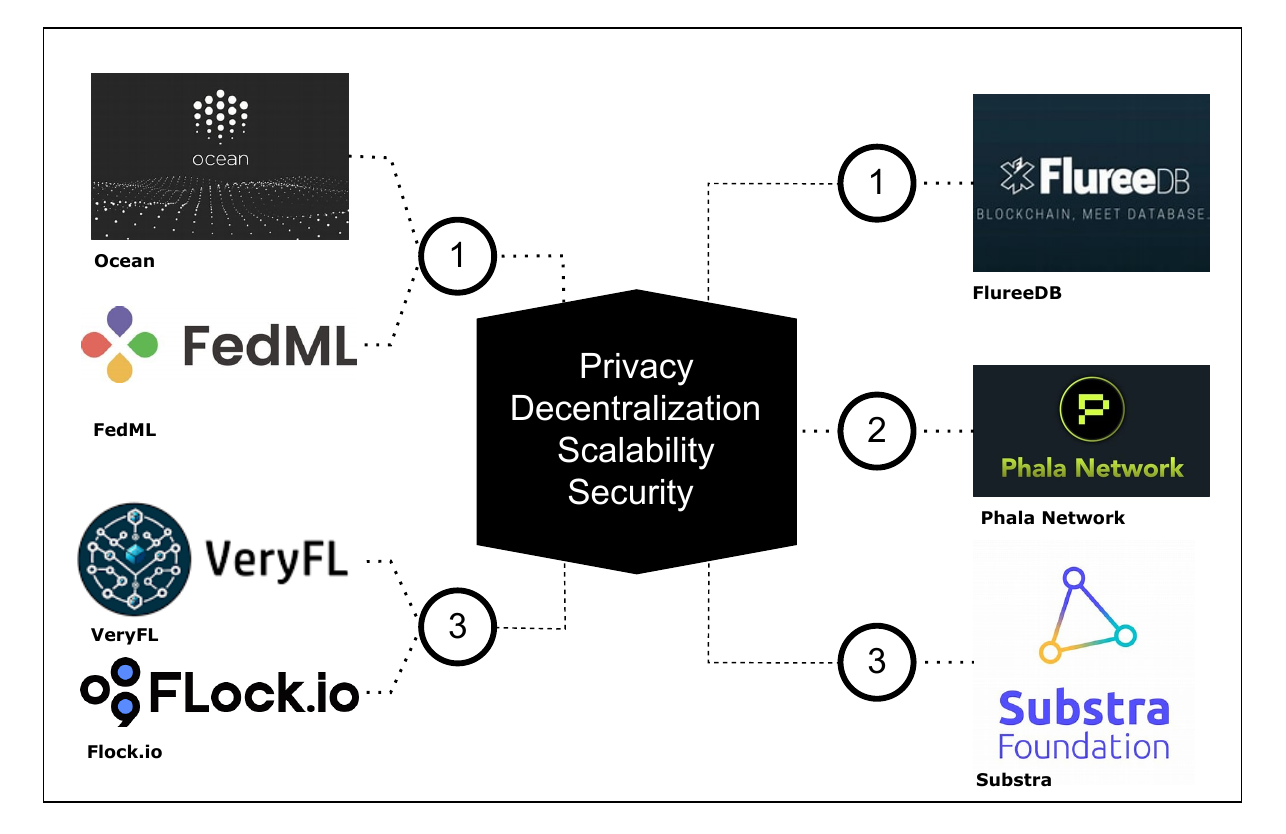}
\caption{The brands depicted in this figure are pioneers in integrating federated analysis with blockchain technology. These projects illustrate the potential for a wide array of applications that merge blockchain with federated analytics, such as data privacy protection, distributed machine learning, decentralized data analysis, and more. This convergence highlights the potential to further augment the role of blockchain technology within the digital economy. In summary, these projects offer practical, real-world applications of blockchain technology.}
\label{fig:ts_bu02}
\end{figure}

Figure \ref{fig:ts_bu02} highlights leading pioneers in the field of integrating federated analysis with blockchain. Our paper aims to contribute to this field by enhancing privacy, scalability, security, and decentralization through the combination of these technologies.

The integration of blockchain and federated analytics can be explored through three key areas:
\begin{itemize}
\item \textbf{Decentralization}: Efficient index calculation is achieved by having each subnet perform data analysis locally under unified rules—such as calculating trading volumes or user activity—and then transmitting only the encrypted results to a central or decentralized coordinator for global index aggregation. This method minimizes data migration, enhancing efficiency and reducing load on the main chain.
\item \textbf{Security}: Federated analytics enhances security and privacy through secure communication protocols, such as multiparty computation (MPC)~\cite{mpc}, ensuring that only necessary, encrypted data is exchanged between subnets. This approach maintains the integrity and privacy of transaction data, aligning with blockchain's transparency and security standards.
\item \textbf{Scalability}: The architecture can dynamically adapt to include new subnetworks or integrate Layer 2 solutions, seamlessly incorporating new data sources and updating protocols as needed. This flexibility supports the ongoing growth and technological evolution of the blockchain ecosystem.
\end{itemize}

Additional opportunities for integrating blockchain techniques with federated analytics include:
\begin{itemize}
\item \textbf{Model Development}: Developing federated analysis models specific to different L2 technologies (like Rollups, side chains, or state channels) to assess and quantify their contributions to metrics such as transaction efficiency, cost, and decentralization levels, thus aiding in technology selection and optimization.
\item \textbf{Cross-Chain Federation Analysis}: Implementing cross-chain federation analysis to build indexes that span multiple blockchain platforms (such as Ethereum, Polkadot, Binance Smart Chain), which can reveal the interactions and collaborative trends across a multi-chain ecosystem, supporting cross-chain interoperability and investment strategies.
\item \textbf{Privacy Enhancement}: Enhancing privacy protection within federated analytics by balancing regulatory compliance needs with robust privacy measures, potentially incorporating technologies like zero-knowledge proofs or trusted execution environments to meet both regulatory and user privacy expectations.
\end{itemize}

This innovative approach not only addresses individual limitations of each system but also unlocks new capabilities by leveraging their mutual strengths in decentralization, security, and efficiency. This promises to significantly advance the state of technology in decentralized finance and beyond. Leading companies are already exploring these synergies, as illustrated in the following graph which highlights major contributors to this integration. Moving forward, challenges such as malicious attacks, resource allocation, and decentralization issues, common to both blockchain and federated analytics, warrant further research~\cite{shi2021federated, dong2024defending}.

A comprehensive evaluation of blockchain performance should integrate both on-chain and off-chain data. Off-chain data, including user sentiment and experience, can provide valuable insights into the broader implications and effectiveness of blockchain systems~\cite{zhang2023mechanics, quan2023decoding, fu2023ai}. 

Additionally, the comprehensive measurement of open-source data and code offers a reliable basis for designing blockchain systems that benefit both technological and human aspects~\cite{liu2022empirical, liu2024economics, fu2023ai, huang2024web3}. This approach also provides essential information for crypto asset investments, contributing to the future of finance~\cite{zhang2022data, yu2024bitcoin, liu2022deciphering, liu2023cryptocurrency}.

\bibliographystyle{ieeetr}
\bibliography{Bibliography}

\appendix


\section{Detailed Definitions of Decentralization Indices}
\footnotesize
\label{subsec: indices}
\textbf{Indice I Adapted $Shannon\ Entropy$}. As entropy is always used to measure the randomness or chaos in a system, the proposed indices aim to measure the degree of randomness in the distribution of controllers. A higher value indicates more chaos in authority distribution, while a lower value refers to a more centralized system. We define the indices $H(v)$ as:
\begin{equation}
    H(v) = \prod \limits_{i=1}^N P(v_i)^{-P(v_i)}
\end{equation}
where the $v_i$ refers to the unit data for each layer and the $P(v_i)$ refers to the weight of the unit data concerning the overall dataset:
\begin{equation}
    P(v_i) = \frac{v_i}{\sum_{i=1}^N v_i}
\end{equation}

\textbf{Indice II $Gini$ $Coefficient$}. As a classical economy indicator, the $Gini$ $Coefficient$ usually indicates the wealth distribution within a given population. Thus, we still consider the $P_i$ as the weight of a unit data concerning the complete dataset and define the indices II as:
\begin{equation}
    G = 1 - \sum_{i=1}^N P_i^2
\end{equation}
a higher indices value indicates less evenness in decentralization distribution, while a lower value shows more decentralization.

\textbf{Indice III $Nakamoto$ $Coefficient$}. The $Nakamoto$ $Coefficient$ is utilized in various scenarios to measure the smallest number of entities that compromise a specific target. For instance, the coefficient is used in Bitcoin analysis to observe the mining power distribution. Here, we suppose that the smallest number of transaction entities or proposer/validator entities to accumulate 51\% of the blockchain can present decentralization in our target layers. Thus, we give the following definition:
\begin{equation}
    N = min\{k \in [1,...,K] : \sum_{i=1}^K P_i > 0.51\}
\end{equation}
where the $P_i$ refers to the weight of a unit of data. In this case, a higher value means better decentralization, for there will be more entities to achieve 51\% 

\textbf{Indice IV $Herfindahl$ $Hirschman$ $Index$}. The $Herfindahl$ $Hirschman$ $Index$ is originally used to measure the market concentration where different firms co-exist. From our perspective, the $HHI$ indices can describe the decentralization for every data unit. Thus, we give the definition:
\begin{equation}
    HHI = \sum_{i=1}^N P_i^2
\end{equation}
where the $P_i$ indicates the share of each data unit concerning the overall dataset. In this case, a lower value refers to more decentralization, while a higher one indicates more centralization.

\section{Detailed Data Dictionary}
\input{tabs/tab5}

\end{document}

%% file: fig/fig0.tex
\begin{figure}
\centering
\begin{tikzpicture}[
    node distance=0.5cm and 0.1cm,
    auto,
    block/.style={rectangle, draw, fill=blue!30, text width=1.6cm, text centered, rounded corners, minimum height=2.8em, font=\scriptsize},
    line/.style={draw, -Latex},
    qblock/.style={rectangle, draw, fill=orange!60, text width=1.6cm, text centered, rounded corners, minimum height=2.8em, font=\scriptsize}
]

\node[block] (method) {Research Methodology: Answering Questions with Empirical Analysis};
\node[qblock, above left=of method, xshift=-1cm]  (q1) {Q1: How to quantify decentralization?};
\node[qblock, above=of method] (q2) {Q2: How to measure scalability?};
\node[qblock, above right=of method, xshift=1cm] (q3) {Q3: How to evaluate security?};
\node[block, below=of method, yshift=-0.5cm] (data) {Data Description \& Solutions};

\node[block, below left=of data, xshift=-1cm] (sol1) {Solution I: Quantify Decentralization with Four Indices};
\node[block, below=of data] (sol2) {Solution II: Measure Scalability with Two Metrics};
\node[block, below right=of data, xshift=1cm] (sol3) {Solution III: Theoretical Analysis of Security};
\node[block, below=3cm of data] (results) {Results: R1, R2, and R3};

\draw[line] (q1) |- (method);
\draw[line] (q2) -- (method);
\draw[line] (q3) |- (method);
\draw[line] (method) -- (data);
\draw[line] (data) -| (sol1);
\draw[line] (data) -- (sol2);
\draw[line] (data) -| (sol3);
\draw[line] (sol1) |- (results);
\draw[line] (sol2) -- (results);
\draw[line] (sol3) |- (results);

\begin{scope}[on background layer]
    \node[draw, dashed, rounded corners, fit=(q1) (q3), inner sep=8pt, fill=yellow!40] {};
    \node[draw, dashed, rounded corners, fit=(sol1) (sol3), inner sep=8pt, fill=yellow!40] {};
\end{scope}

\end{tikzpicture}
\caption{Overview of the research structure and methodologies.}
\label{fig:flowchart}
\end{figure}

%% file: tabs/tab1.tex
\begin{table*} 
\caption{Data Form for Ethereum 2.0}
\label{table:1}
\centering
\begin{tabularx}{0.8\textwidth}{@{}lXl@{}}
\toprule
Data Type   & Data Frame                                                                                                  & Description                                                                                                                                           \\ 
\midrule
Block       & \begin{tabular}[c]{@{}l@{}}Daily Block Count\\ Average Block Time\\ Average Gas Used by Blocks\end{tabular} & \begin{tabular}[c]{@{}l@{}}Number of blocks produced per day\\ Average consensus time per block\\ Average gas used per block\end{tabular}             \\ 
\midrule
Transaction & \begin{tabular}[c]{@{}l@{}}Transaction Count\\ Gas Limit\\ Burned Fees\end{tabular}                         & \begin{tabular}[c]{@{}l@{}}Transaction count per day\\ Gas limit amount per day\\ Used tokens for transactions per day\end{tabular}                     \\ 
\midrule
Account     & \begin{tabular}[c]{@{}l@{}}Validator Count\\ Average Validator Balance\\ Participation Rate\end{tabular}    & \begin{tabular}[c]{@{}l@{}}Validator counts per day\\ Average account balance of validators per day\\ Overall participation rate per day\end{tabular} \\ 
\midrule
Network     & Network Liveness                                                                                            & Block count for confirmation                                                                                                                          \\ 
\bottomrule
\end{tabularx}
\end{table*}

%% file: tabs/tab2.tex
\begin{table*} 
\caption{Data Form for Algorand}
\label{table:2}
\centering
\begin{tabularx}{0.8\textwidth}{@{}lXl@{}}
\toprule
Data Type   & Data Frame                                                              & Description                                                                                       \\ 
\midrule
Block       & \begin{tabular}[c]{@{}l@{}}Block Info\\ Proposer Count\end{tabular}     & \begin{tabular}[c]{@{}l@{}}Block timestamp, address, height\\ Proposer count per day\end{tabular} \\ 
\midrule
Transaction & \begin{tabular}[c]{@{}l@{}}Transaction Count\\ Burned Fees\end{tabular} & \begin{tabular}[c]{@{}l@{}}Transaction count per day\\ Tokens used for transactions\end{tabular}   \\ 
\midrule
Account     & Block Reward                                                            & Reward for block proposal per day                                                                 \\ 
\midrule
Contract    & \begin{tabular}[c]{@{}l@{}}Contract Calls\\ Unique Calls\end{tabular}   & \begin{tabular}[c]{@{}l@{}}Overall contract calls per day\\ Unique contract calls\end{tabular}    \\ 
\bottomrule
\end{tabularx}
\end{table*}

%% file: tabs/tab3.tex
\begin{table*}[!htbp]
\caption{The Decentralization Indices for Layers}
\label{table:dec_result}
\begin{tabularx}{\textwidth}{X *{4}{>{\centering\arraybackslash}X}}
\toprule
\multirow{2}{*}{Blockchain} & \multicolumn{2}{c}{Consensus Layer} & \multicolumn{2}{c}{Transaction Layer} \\
\cmidrule(lr){2-3} \cmidrule(lr){4-5}
& Indices & Values & Indices & Values \\
\midrule
\multirow{4}{*}{Algorand} 
& Shannon Entropy & 1364.34 & Shannon Entropy & 920.192 \\
& Gini Coefficient & 0.155 & Gini Coefficient & 0.155 \\
& Nakamoto Coefficient & 821 & Nakamoto Coefficient & 931 \\
& Herfindahl Hirschman Index & 0.0005 & Herfindahl Hirschman Index & 0.00015 \\
\midrule
\multirow{4}{*}{Ethereum 2.0} 
& Shannon Entropy & 866.759 & Shannon Entropy & 2252.60 \\
& Gini Coefficient & 0.301 & Gini Coefficient & 0.301 \\
& Nakamoto Coefficient & 705 & Nakamoto Coefficient & 2067 \\
& Herfindahl Hirschman Index & 0.0021 & Herfindahl Hirschman Index & 0.0004 \\
\bottomrule
\end{tabularx}
\end{table*}

%% file: tabs/tab5.tex
\definecolor{col1}{RGB}{229,243,255} 
\definecolor{col2}{RGB}{204,229,255}
\definecolor{col3}{RGB}{179,217,255}
\definecolor{col4}{RGB}{153,204,255}
\definecolor{col5}{RGB}{128,191,255}
\definecolor{col6}{RGB}{102,178,255}
\definecolor{col7}{RGB}{77,166,255}
\definecolor{col8}{RGB}{51,153,255}

\begin{table}[h]
\caption{ On-chain data for Algorand from January 2019 to September 2023 via BitQuery’s open APIs, and for Ethereum 2.0 from June 2019 to September 2023 through Beacon Explorer using the SPIDER framework.}
\label{table:5}

\begin{minipage}[t]{0.5\textwidth}
\vspace{0pt} 
\flushright
\small
\resizebox{0.98\textwidth}{!}{
\begin{tabular}{%
    >{\columncolor{col1}}p{1.4cm} 
    >{\columncolor{col2}}p{1.9cm}
    >{\columncolor{col3}}p{2cm} 
    >{\columncolor{col4}}p{0.5cm} 
    >{\columncolor{col5}}p{0.9cm} 
    >{\columncolor{col6}}p{1.3cm} 
    >{\columncolor{col7}\color{black}}p{3.3cm} 
    >{\columncolor{col8}\color{black}}p{5.0cm}
}
\toprule
\textbf{Data Type} & \textbf{Data Frame} & \textbf{Description} & \textbf{Unit} & \textbf{Type} & \textbf{Frequency} & \textbf{Range} & \textbf{File Name} \\ 
\midrule
Block & Daily Block Count & Numbers of blocks produced per day & NA & Integer & Daily & 0\textasciitilde{}7180 & daily\_block\_count.csv \\
 & Average Block Time & Average consensus time per block & S & Float & Daily & 4.46\textasciitilde{}30.57 & avg\_blk\_time.csv \\
 & Average Gas Used by Blocks & Average gas used per block & NA & Float & Daily Sum & 0\textasciitilde{}15511762.25 & gas\_used\_avg\_by\_blk.csv \\
Transaction & Transaction Count & Transaction count per day & NA & Integer & Daily & 0\textasciitilde{}1932226 & daily\_transactions.csv \\
 & Gas Limit & Gas limit amount per day & Eth & Integer & Daily Sum & 5000\textasciitilde{}30076713.92 & gas\_limit.csv \\
 & Burned Fees & Used tokens for transactions per day & Eth & Float & Daily & 0\textasciitilde{}71718.88 & burned\_fees.csv \\
Account & Validator Count & Validator counts per day & NA & Integer & Daily & 21063\textasciitilde{}771738 & validator\_data.csv \\
 & Average Validator Balance & Average account balance of validators per day & Eth & Float & Daily & 32.00953203\textasciitilde{}34.00950871 & validator\_avg\_balance.csv \\
 & Participation Rate & Overall participation rate per day & NA & Float & Daily(\%) & 0.941524213\textasciitilde{}0.99728444 & participation\_rate.csv \\
Network & Network Liveness & Block count for confirmation & NA & Integer & Daily & 2\textasciitilde{}12 & network\_Liveness.csv \\
Block & Block Info & Block timestamp, address, height & NA & String & Daily & NA & al\_block\_data.csv \\
 & Proposer Count & Proposer count per day & NA & Integer & Daily & 31\textasciitilde{}130 & al\_block\_data\_proposercount\_reward.csv \\
Transaction & Transaction Count & Transaction count per day & NA & Float & Daily & 913\textasciitilde{}9271981 & al\_transac\_data\_count\_fee.csv \\
 & Burned Fees & Tokens used for transactions & Algo & Float & Daily & 1.47588\textasciitilde{}33113.44687 & al\_transac\_data\_count\_fee.csv \\
Account & Block Reward & Reward for block proposal per day & Algo & Float & Daily & 141.059024\textasciitilde{}5184.994864 & al\_block\_data\_reward.csv \\
Contract & Contract Calls & Overall contract calls per day & NA & Integer & Daily & 1\textasciitilde{}197459 & al\_contracts\_calls\_unique\_calls.csv \\
 & Unique Calls & Unique contract calls & NA & Integer & Daily & 1\textasciitilde{}10149 & al\_contracts\_calls\_unique\_calls.csv \\
\bottomrule
\end{tabular}%
}
\end{minipage}

\label{tab: dictionary}
\end{table}